\documentclass[]{ws-procs9x6square}
\usepackage{float}
\usepackage{epsfig}

\newcommand{\MSO}{MSO}%{marginally stable orbit}
\newcommand{\ms}{{\ensuremath{\rm ms}}}
\newcommand{\up}[1][{}]{\ensuremath{u^{#1}_{\rm \hat p}}}
\newcommand{\cs}[1][{}]{\ensuremath{c^{#1}_{\rm s}}}
\newcommand{\ie}{\textit{i.e.}}
\newcommand{\eg}{\textit{e.g.}}
\newcommand{\etal}{\textit{et al.}}
\newcommand{\cf}{\textit{cf.}}

\newcommand{\rms}[1][{}]{\ensuremath{r^{#1}_{\rm ms}}}

\newcommand{\Thms}[1][{}]{\ensuremath{\Theta^{#1}_{\rm ms}}}

\title{Two-dimensional structure of thin transonic discs: observational
manifestations}

\author{V.S. Beskin}
\address{Lebedev Physical Institute,\\
Leninskii prosp., 53, Moscow, 119991, Russia.\\
E-mail: beskin@lpi.ru}
\author{A.D. Tchekhovskoy}
\address{Moscow Institute of Physics and Technology,\\
Institutskii per., 9, Dolgoprudny, 141700, Russia.\\E-mail:
chekhovs@lpi.ru}

\begin{document}

\maketitle

\abstracts{We study the two-dimensional structure of thin
transonic accretion discs in the vicinity of a non-spinning black
hole within the framework of hydrodynamical version of the
Grad-Shafranov equation. Our analysis focuses on the region inside
the marginally stable orbit (\MSO), $r < \rms$. We show that all
components of the dynamical force in the disc become significant
near the sonic surface and (especially) in the supersonic region.
Under certain conditions, the disc structure can be far from
radial, and we review the affected disc properties, in particular
the role of the critical condition at the sonic surface. Finally,
we present a simple model aimed at explaining the quasi-periodical
oscillations that have been observed
in the infra-red and X-ray radiation of the Galactic Centre.%
}

\section{Introduction}
\label{Sec.intro} The investigation of accretion flows near black
holes (BHs) is undoubtedly of great astrophysical interest.
Substantial energy release must take place near BHs, and general
relativity effects, attributable to strong gravitational fields,
must show up there. Depending on external conditions, both
quasi-spherical and disc accretion flows can be realized. The
structure of thin accretion discs has been the subject of many
papers. Many results were included in
textbooks~\cite{sha83,lip92}. Lynden-Bell \cite{lyn69} was the
first to point out that supermassive BHs surrounded by accretion
discs could exist in galactic nuclei. Subsequently, a theory for
such discs was developed that is now called the standard model, or
the model of the $\alpha-$disc~\cite{sha72,sha73,nov73}.

Since then the standard disc thickness prescription has been
widely used,
\begin{equation}
H \approx r\frac{\cs}{v_{\rm K}}, \label{2}
\end{equation}
where the disc thickness $H$ is assumed to be determined by the
balance of gravitational and accreting matter pressure forces with
the dynamical force neglected. This relation was later used in the
renowned approach where all quantities were averaged over the disc
thickness~\cite{pac81}, with a lot of such one-dimensional models
following~\cite{abr88,pap94,rif95,che97,nar97,pei97,bel98,gam98a,gam98b,art01}.
As for the two-dimensional structure of accretion discs, it was
investigated mostly only nume\-rical\-ly and on\-ly for thick
discs~\cite{pap94,igu97,bal98,kro02}.

Even though standard disc thickness prescription \eqref{2} and the
averaging procedure are likely to be valid in the region of stable
orbits \hbox{$r > \rms$}~\cite{sha83}, they in our view require a
more serious analysis. It is the assumption that the transverse
velocity $v_\theta$ may \emph{always} be neglected in thin
accretion discs all the way up to horizon~\cite{abr97}, \ie\ that
the disc thickness is always determined by (\ref{2}), that is the
most debatable~\cite{bes04}. This assumption is widely used,
explicitly or implicitly, virtually in all papers devoted to thin
accretion discs~\cite{abr81,cha96}.

There is a brief discussion on thin discs in this volume that
covers theory shortcomings and a brief description of our approach
(\cf\ Sec.\ named Thin disk in~\cite{bes04}). In this paper we
briefly describe our study of subsonic and transonic regions of
thin discs followed in Sec.\hbox{}~\ref{Sec.SupersonicFlow} by the
elaborate discussion of the supersonic flow. Finally, in
Sec.\hbox{}~\ref{Section.Observ} we develop a toy model for
explaining the observed quasi-periodical oscillations detected in
the infra-red and X-ray observations of the GC.

\section[]{Basic equations}
\label{Sec.BasicEqs} We consider thin disc accretion on to a BH in
the region where there are no stable circular orbits. The
contribution of viscosity should no longer be significant
here~\cite{bes04}. Hence we may assume that an ideal hydrodynamics
approach is suitable well enough for describing the flow structure
in this inner area of the accretion disc. Below, unless
specifically stated, we consider the case of non-spinning BH, \ie\
use the Schwarzschild metric, and use a system of units with
$c=G=1.$ We measure radial distances in the units of $M$, the BH
mass.

In \hbox{Boyer-Lindquist} coordinates the Schwar\-z\-schild metric
is~\cite{lan87a}
\begin{equation}
 {\rm d}s^{2}=-\alpha^{2}{\rm d}t^{2}
 + g_{ik}{\rm d}x^{i}{\rm  d}x^{k}, \label{a1}
\end{equation}
where
\begin{gather}
 \alpha^2 = 1-2/r, \quad g_{rr} = \alpha^{-2}, \quad
 g_{\theta \theta} = r^2, \quad
 g_{\varphi \varphi} = \varpi^{2} =
 r^2\sin^2\theta.\label{a2}
\end{gather}

We reduce our discussion to the case of axisymmetric stationary
flows. For an ideal flow there are three integrals of motion
conserved along the streamlines, namely entropy, $S$, energy  $E =
\mu\alpha\gamma$, and $z$-component of angular momentum $L =
\mu\varpi u_{\hat\varphi}$, where $\mu=(\rho_{m}+P)/n$ ($\rho_{m}$
is internal energy density, $P = n T$ is pressure) is relativistic
enthalpy. The relativistic Bernoulli equation $\up[2] =
\gamma^2-u_{\hat\varphi}^2 - 1$, where \up\ is the physical
poloidal $4$-velocity component~\cite{bes04,bes97,bes02}, now
becomes
\begin{align}
\up[2] &= \frac{E^2-\alpha^2L^2/\varpi^2
-\alpha^2\mu^2}{\alpha^2\mu^2}. \label{up2full}
\end{align}

Below we use another angular variable $\Theta = \pi/2 - \theta$
and for the sake of simplification we adopt the polytropic
equation of state $P = k(S)n^{\Gamma}$ so that temperature and
sound velocity can be written as~\cite{sha83}
\begin{equation}
T = k(S)n^{\Gamma - 1}; \qquad \cs[2] =
\frac{\Gamma}{\mu}k(S)n^{\Gamma - 1}. \label{Tc}
\end{equation}

\section[]{Subsonic flow}
\label{Sec.SubsonicFlow}

Following Sec.\hbox{}~\ref{Sec.intro}, we assume that the
$\alpha$-disc theory holds outside the \MSO. We adopt the flow
velocity components, which this theory yields on the \MSO\
$r=\rms,$\footnote{A nearly parallel inflow with a small radial
velocity $v_r \approx \alpha_{\rm SS} \cs[2]/v_{\rm K} \ll \cs \ll
1$; $\rms = 3 r_{\rm g}$, where $r_{\rm g} = 2M$ is the
gravitational radius of the BH of mass $M$.} as the first three
boundary conditions for our problem. For the sake of simplicity we
consider the radial velocity, which is responsible for the inflow,
to be constant at the surface $r = \rms$ and equal to $u_0$ and
the toroidal velocity to be exactly equal to that of a free
particle revolving at $r=\rms$.%
\footnote
 {
 For a free particle revolving at $r=\rms$ around a non-spinning
 BH we have \cite{lan87a} $u_{\hat\varphi}(\rms) = 1/\sqrt{3},
 \alpha_0 = \alpha(\rms) = \sqrt{2/3}, \gamma_0 = \gamma(\rms) =
 \sqrt{4/3}.$ \label{Ftn.MS}
 }
We also assume the speed of sound to be constant at the \MSO, $\cs
= c_0 = \mathrm{const}$. Having introduced $\Thms$ \cite{bes02}
--- the Lagrange coordinate of streamlines at the \MSO\ --- for
$\cs \ll 1$, \ie\ non-relativistic temperature, we obtain from
(\ref{up2full}) and (\ref{Tc}),
\begin{equation}
\up[2]  = u_0^2 +w^2 + \frac{2}{\Gamma - 1}\left(c_0^2 -
\cs[2]\right) + \frac{1}{3}\left(\Thms[2]-\Theta^2\right) + \dots
\label{uup}
\end{equation}
The quantity
\begin{equation}
w^2(r) = \frac{e_0^2-\alpha^2l_0^2/r^2-\alpha^2}{\alpha^2} \equiv
\frac{1}{\alpha^2}\frac{(6-r)^3}{9r^3}, \label{w2}
\end{equation}
where $e_0 = E_0/\mu_0$ and $l_0 = L/(\mu_0 \cos \Thms)$, is the
poloidal four-velocity of a free particle having zero poloidal
velocity at the \MSO.

In the extreme subsonic case, $\up \ll \cs$, the Grad-Shafranov
hydrodynamic equation is significantly
simplified~\cite{bes04,bes02}. The numerical results are shown in
Fig.\hbox{}~\ref{Fig.global}. In the subsonic region, $r_* < r \le
\rms \equiv 3 r_{\rm g}$, the disc thickness rapidly diminishes,
and at the sonic surface we have $H(r_*) = u_0/c_0 H(\rms)$, so
that you cannot neglect the dynamical force there.

We stress that taking the dynamical force into account is indeed
extremely important. This is because, unlike zero-order standard
disc thickness prescription \eqref{2}, the Grad-Shafranov equation
has second order derivatives, \ie\ contains two additional degrees
of freedom. This means that the critical condition only fixes one
of these degrees of freedom (\eg\ imposes some limitations on the
form of the flow) rather than determines the angular momentum of
the accreting matter~\cite{bes04,bes02}.

\section[]{Transonic flow}
\label{Sec.TransFlow}

In order to verify our conclusions we consider the flow structure
in the vicinity of the sonic surface in more detail. Since the
smooth transonic flow is analytical at a singular point $r = r_*,
\Theta = 0$~\cite{lan87b}, it is possible to express the
quantities via a series of powers of $h = (r-r_*)/r_*$ and
$\Theta$. Substituting these expansions into equations of motion
we get a set of equations on the coefficients which allows us to
reconstruct the flow structure in the vicinity of the sonic point
(see Fig.\hbox{}~\ref{Fig.transonic}), in particular
\begin{align}
 \up[2] &= c_*^2
   \left[1-2\eta_1 h+
         \frac{1}{6}(\Gamma - 1) \, \frac{a_0^2}{c_0^2}\Theta^2
                  + \frac{2}{3}(\Gamma + 1)\eta_1^2\Theta^2
   \right], \nonumber \\
 \cs[2] &= c_*^2
   \biggl[1+\left(\Gamma-1\right)\eta_1 h +
        \frac{1}{6}(\Gamma -1 ) \, \frac{a_0^2}{c_0^2}\Theta^2
       - \frac{1}{3}(\Gamma - 1)(\Gamma+1) \eta_1^2 \Theta^2
   \biggr], \label{cs}
\end{align}
where $a_0 = \left[2/(\Gamma+1)\right]^
{(\Gamma+1)/2(\Gamma-1)}c_0/u_0$ gives the compression of
streamlines, $a_0 = H(\rms)/H(r_*)$, and $\eta_1 \sim u_0^{-1}$.
Equation \eqref{cs} yields shape of the sonic surface, $\up =
\cs$; it has the standard parabolic form $h =
(\Gamma+1)\eta_1\Theta^2/3$.

Since the transonic flow in the form of a nozzle (see
Fig.\hbox{}~\ref{Fig.transonic}) has
longitudinal and transversal scales of one order of
magnitude~\cite{lan87b},
 near the sonic surface we have $\delta r_\parallel
\approx \delta r_\bot,$ \ie\ $\delta r_\parallel \approx H(r_*)$.
Hence for thin discs (\ie\ for $c_0 \ll 1$) this longitudinal
scale is always much smaller than the distance from the BH,
$\delta r_\parallel/r_* \approx H(r_*)/r_* \ll 1$. Only by taking
the transversal velocity into account do we retain the small
longitudinal scale $\delta r_\parallel \ll r_{\rm g}.$ This scale
is left out during the standard one-dimensional approach.

\section[]{Supersonic flow}
\label{Sec.SupersonicFlow}
Since the pressure gradient becomes
insignificant in the supersonic region, the matter moves here
along the trajectories of free particles. Neglecting the
$\nabla_{\theta}P$ term in the $\theta$-component of relativistic
Euler equation~\cite{fro98}, we have~\cite{abr97}
\begin{equation}
 \alpha u_{\hat r} \frac{\partial (r u_{\hat \Theta})} {\partial r}
 + \frac{(r u_{\hat \Theta})}{r^2}
 \frac{\partial (r u_{\hat \Theta})} {\partial \Theta} +
 (u_{\hat\varphi})^2 \tan\Theta = 0.
\label{theuler}
\end{equation}
Here, using the conservation law of angular momentum,
$u_{\hat\varphi}$ can be easily expressed in terms of radius:
$u_{\hat\varphi} = 2 \sqrt{3}/r$. We also introduce dimensionless
functions $f(r)$ and $g(r)$: $\Theta f(r) = r u_{\hat \Theta}$ and
$g(r) = -\alpha u_{\hat r} > 0$. Using (\ref{theuler}) and the
definitions above, we obtain an ordinary differential equation for
$f(r)$ which could be solved if we knew $g(r)$:
\begin{equation}
\frac{{\rm d} f}{{\rm d} r} = \frac{f^2 + 12}{r^2 g(r)}.
\label{Eq.f}
\end{equation}

From (\ref{up2full}) we have \hbox{$\up[2] \rightarrow w^2$} as
\hbox{$r \rightarrow r_{\rm g}$}. On the other hand, $\up \approx
c_* \approx c_0$ for $r \lesssim r_*$. Therefore, the following
approximation should be valid throughout the \hbox{$r_{\rm g} < r
< r_*$} region, $g(r) \approx \sqrt{(\alpha w)^2 + (\alpha
c_*)^2}$.

Equation \eqref{Eq.f} governs the supersonic flow structure for
the case of non-spinning BH. To get a better match with
observations (\cf\hbox{} Sec.\hbox{} \ref{Section.Observ}), we
also consider a more general case of spinning BH, \ie\hbox{} a
Kerr BH with non-zero specific angular momentum $a$. After some
calculation, equation \eqref{Eq.f} can be generalized to the Kerr
metric with a strikingly simple form,
\begin{equation}
 \frac{{\rm d}f}{{\rm d}r} = \frac{f^2 + a^2
 \left(1-e_0^2\right)+l_0^2}{r^2 \tilde{g}\left(r\right)},
 \label{Eq.fKerr}
\end{equation}
where $\tilde{g}\left(r\right)$ is a straightforward
generalization of $g(r)$ to the Kerr case; we omit it here due to
space limitations. For the Schwarzschild BH ($a=0$,
$e_0=\sqrt{8/9}$, $l_0 = 2\sqrt{3}$ \cite{lan87b}) equation
\eqref{Eq.fKerr} reduces back to \eqref{Eq.f}.

Integrating (\ref{Eq.fKerr}), we obtain
\begin{equation}
f\left(r\right) = \kappa \tan \left[
               \kappa \int_{r_*}^r \frac{d\xi}{\xi^{2}g(\xi)} +
                \frac{\pi}{2}
     \right],
\end{equation}
where $\kappa = \sqrt{a^2\left(1-e_0^2\right)+l_0^2}$ and $\pi/2$
has been to a good accuracy substituted for the integration
constant $\arctan \left[f(r_*)/\sqrt{3}\right]$.\footnote{For $r$
just below $r_*,$ the function $f$ should be positive to reflect
the fact that the flow diverges. Then, $f=0$ corresponds to the
point where the divergency finishes, and the flow starts to
converge.}

\begin{figure}%[htb]
\begin{minipage}[b]{0.65\linewidth}
    \begin{center}
    \epsfig{figure=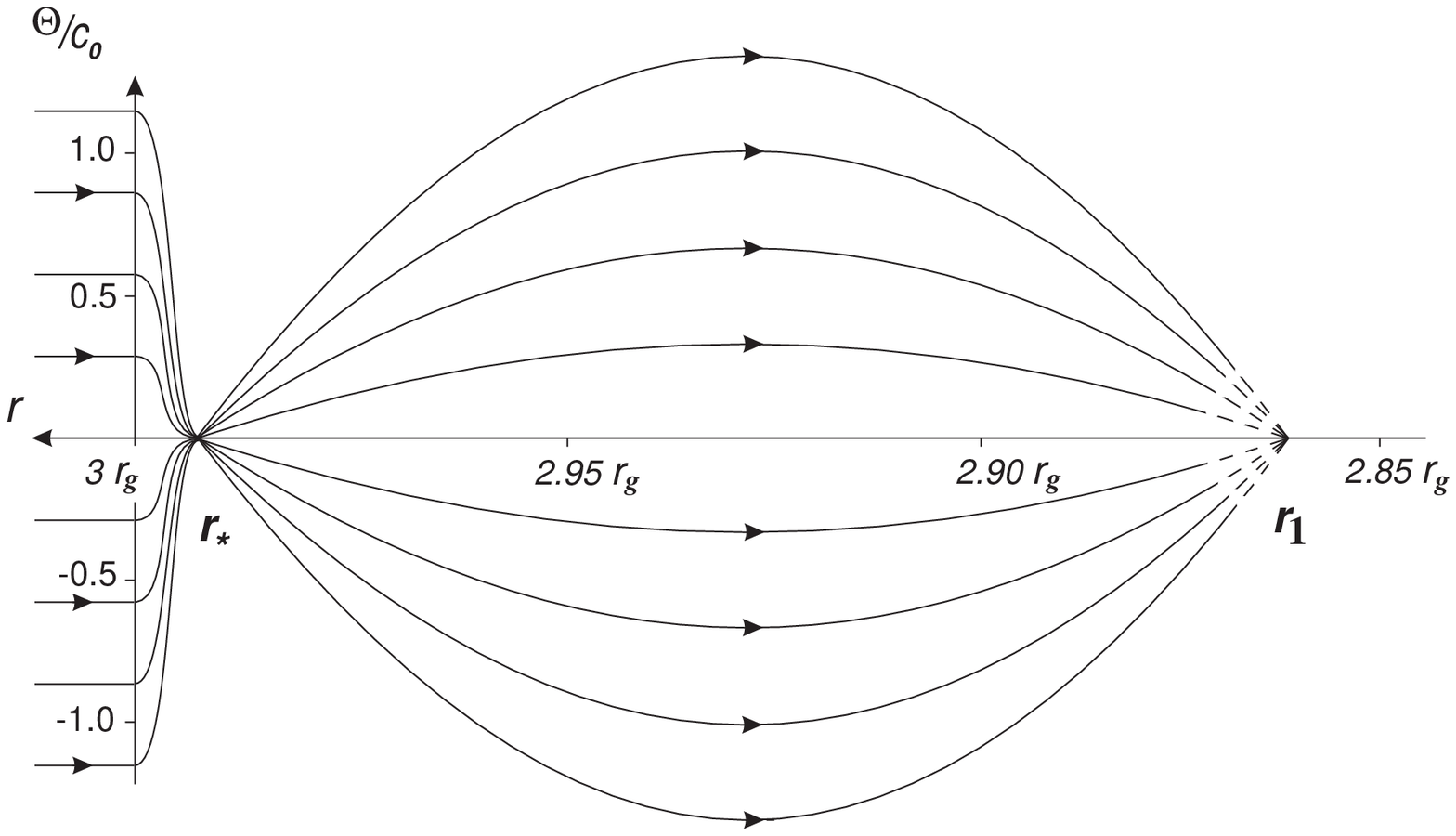,width=\linewidth}
    \caption{The structure of a thin accretion
         disc (actual scale) for $c_0 = 10^{-2}$, $u_0 = 10^{-5}$
         after passing the \MSO\ $r = 3r_{\rm g}$ ($a=0$, Schwarzschild case).
         As sufficient dissipation
         can take place in the vicinity of the first node $r = r_{\rm 1}$,
         we do not prolong the flow lines to the region $r < r_{\rm 1}$.}
     \label{Fig.global}
     \end{center}
\end{minipage}\hfill
\begin{minipage}[b]{0.3\linewidth}
  \begin{center}
    \epsfig{figure=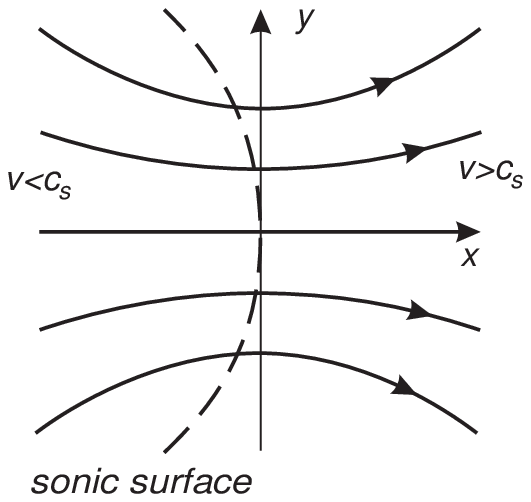,width=\linewidth}
    \caption{Schematics of thin disc stre\-am\-lin\-es pro\-file around the
    sonic point. The flow has the form of the standard nozzle.
    Here \hbox{$x = -h$}, and \hbox{$y = \Theta$}.}
     \label{Fig.transonic}
   \end{center}
\end{minipage}
\end{figure}

The results of numerical calculations are presented in
Fig.\hbox{}~\ref{Fig.global}.
In the supersonic region the flow
performs transversal oscillations about the equatorial plane,
their frequency independent of their amplitude. We see as well
that the maximum thickness of the disc in the supersonic (and,
hence, ballistic) region, which is controlled by the transverse
component of the gravitational force, actually coincides with the
disc thickness within the stable orbits region, $r > \rms$, where
standard estimate (\ref{2}) is correct.

\begin{figure}
\begin{minipage}[b]{0.64\linewidth}
  \begin{center}
    \epsfig{figure=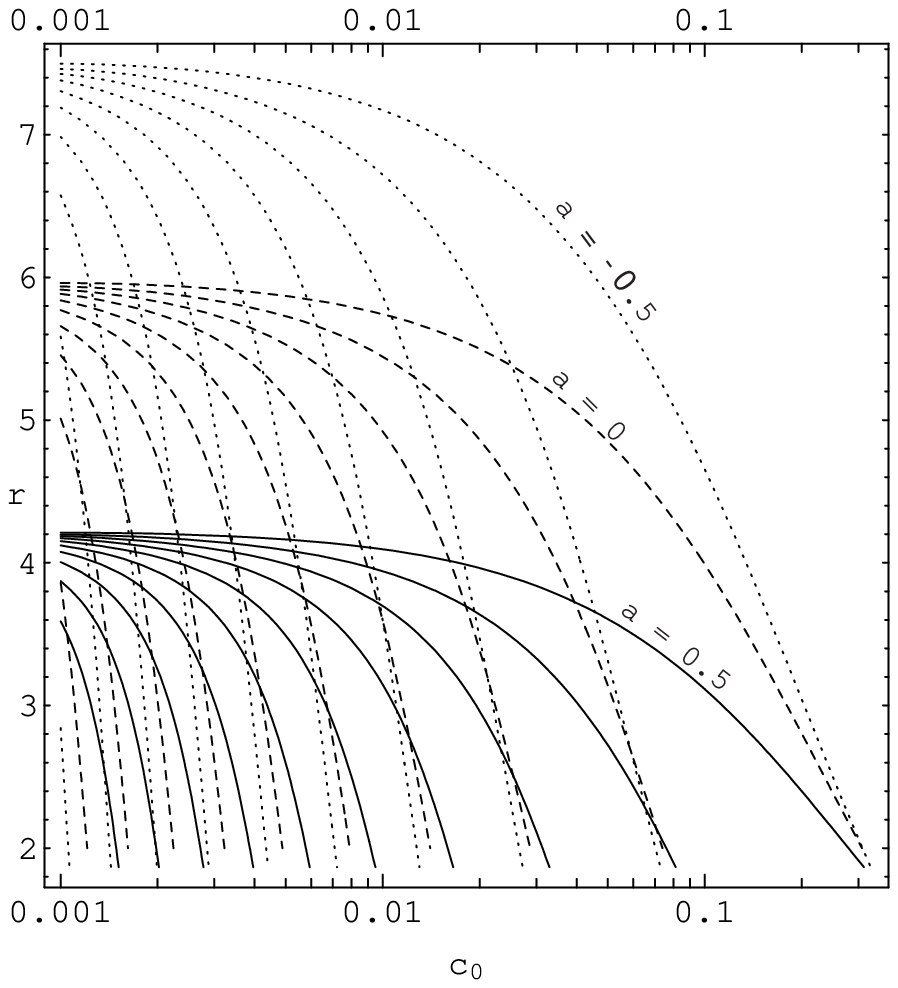,width=\linewidth}
    \caption{Radial positions (in the units of $M$) of the nodes
    for a range of initial sound velocities.
    Dotted, dashed, and solid curves correspond to the cases
    $a=-0.5$, $a=0$, and $a=0.5$ respectively.
    Each curve relates the radial position of a
    node to a value of the initial sound velocity. Intersection
    points of these curves with the line $c_0 = {\rm const}$ give the
    the nodes' radial positions for that particular value of $c_0$.}
     \label{Fig.nodespos}
   \end{center}
\end{minipage}\hfill
\begin{minipage}[b]{0.32\linewidth}
  \begin{center}
    \epsfig{figure=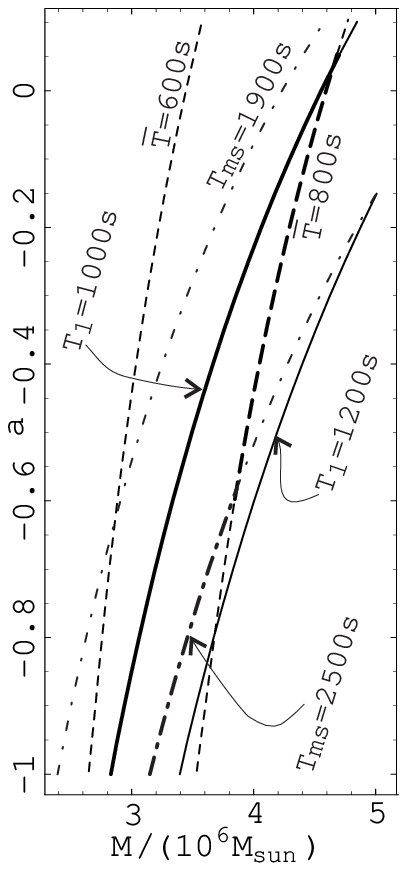,width=\linewidth}
    \caption{Relation of $a$ vs.\ $M$.
    Dashed, dash-dotted, and solid lines come from matching
    $\overline T$ ($700 \pm 100$ s), $T_\ms$ ($2200 \pm 300$
    s), and $T_0$ ($1100 \pm 100$ s) respectively.
    The resulting error polygon is bolded.}
    \label{Fig.aM}
  \end{center}
\end{minipage}
\end{figure}

Once diverged, the flow converges once again at a `nodal' point
closer to the BH. The radial positions the nodes are given by the
implicit formula $f(r_n)=\pm\infty,$ \ie\
\begin{equation}
\kappa
\int_{r_n}^{r_*} \frac{d\xi}{\xi^{2}g(\xi)} = n \pi,
\end{equation}
where $n$ is the node number; the node with $n=0$ corresponds to
the sonic surface. In this formula the sonic radius $r_* \equiv
r_0$ can be to a good accuracy approximated by $\rms = \rms(a)$
the expression for which can be found in most
textbooks~\cite{sha83}. Figure~\ref{Fig.nodespos} shows the
positions of nodes for different values of $c_0$ (the positions do
not depend on $u_0$ for $u_0 \ll c_0$) and the BH spin parameter
$a$.
The matter travel time between the nodes has weak dependence not
only on $u_0$ but also on $c_0$ as well. This provides a means for
testing the theory via observations, and we do this in the
following section.

\section{Applications to observations}
\label{Section.Observ}
Suppose some perturbation in the disc (a
``chunk'') approaches the \MSO. We expect to observe radiation
coming from the chunk with the period of its orbital motion,
\begin{equation}
 T_{\rm ms}\left(a\right) = 2\pi
            \left(\rms[3/2]+a\right),
 \label{Eq.Tcirc}
\end{equation}
where $a$ is the angular momentum per unit mass of the BH and
\rms\ is an estimate of the distance from the BH to the
chunk~\cite{sha83}. After a number of rotations, the chunk reaches
the \MSO\ and passes through the nodal structure derived earlier
(\cf\hbox{} Sec.\hbox{}~\ref{Sec.SupersonicFlow}) generating a
flare. Each time the chunk passes through a node, it generates
some additional radiation, and therefore the flare is likely to
consist of several peaks. We believe that it is these peaks that
were discovered in the infra-red and X-ray observations of the
GC~\cite{asc04,gen03}.

The time interval between the detection of two subsequent peaks
equals the time it takes for the chunk to pass between two
adjacent nodes (\hbox{$n$-th} and \hbox{$(n-1)$-th}), $T_n^{(1)}$,
plus the difference in travel times to the observer for the
radiation coming from the $n$-th and $(n-1)$-th nodes,
$T_n^{(2)}$:
\begin{equation}
T_n = T_n^{(1)} + T_n^{(2)}, \label{Eq.T}
\end{equation}
where $n$ is the index of the observed time interval (counting
from one).

The first term in r.h.s.\ of \eqref{Eq.T} can be easily obtained
from the analysis of particle's geodesics in the equatorial
plane~\cite{sha83}
\begin{equation}
 T_n^{(1)} \left(a, \cs\right)=
  \int_{r_n}^{r_{n-1}}
    \frac{u^t}{u^r}
    {\rm d}r =
  \int_{r_n}^{r_{n-1}}
    \frac{\left(-e_0 g^{tt} +l_0 g^{t\varphi}\right)|_{\theta=\pi/2}}
    {g\left(r\right)} {\rm d}r,
\end{equation}
where the coefficients of the inverse metric are $g^{tt} = -
\Sigma^2/(\rho^2 \Delta)$ and $g^{t\varphi} = \omega g^{tt}$; the
definitions of $\Sigma$, $\rho$, $\Delta$, and $\omega$ can be
found elsewhere in this volume~\cite{bes04}.

\begin{figure}[bth]
  \begin{center}
    \includegraphics[height=6.5cm, keepaspectratio]{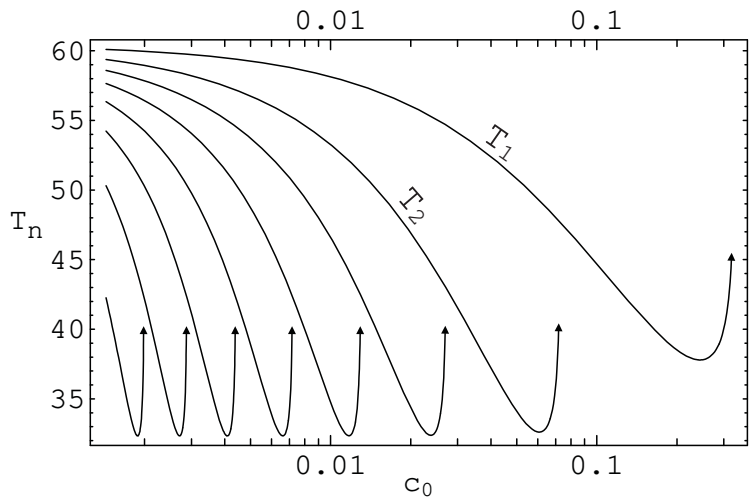}
     \caption{
      The dependence of time intervals between the peaks in a flare
      on the speed of sound in the disc, $c_0$, for a moderately spinning BH
      ($a=-0.5$). The uppermost
      curve corresponds to the time interval $T_1$ between the
      the $0$th and $1$st peaks,
      the second curve
      from the top corresponds to the time interval $T_2$ between the $1$st and $2$nd ones,
      etc.
      All intervals $T_n$ behave very similarly: they first
      decrease with $c_0$
      and then abruptly increase to infinity due to time dilation when
      the innermost node in the pair comes close to the BH horizon (which is
      indicated with upward arrows).
      Even though individual time intervals between subsequent peaks in
      the flare may depend on the temperature in the disc (which is proportional
      to $c_0^2$, see \eqref{Tc}), their minimum and maximum values remain the
      same for the range of sound velocities where there are several intervals observed.
      In the particular case of $a=-0.5$, illustrated in the figure, we have
      $T_{\rm min} \approx 32$ and $T_{\rm max} \approx 60$
      with all other time intervals lying in between.
      }
      \label{Fig.periods}
   \end{center}
\end{figure}
For definiteness and simplicity, we assume that the observer is
located along the rotation axis of the BH. On its way to the
observer, the radiation travels along the null geodesic that
originates at a node in the equatorial plane (\eg\ $r=r_n$,
$\theta = \pi/2$% for the $n$-th node
) and reaches the observer at infinity ($r = \infty$, $\theta =
0$). Using these as boundary conditions for null geodesics in the
Kerr metric~\cite{car68}, we numerically find $T_n^{(2)}(a,\cs)$.

Figure \ref{Fig.periods} shows the dependence of observed time
intervals on the value of the speed of sound in the disc. Although
each individual time interval may depend on $c_0$, the range
[$T_{\rm min}(a)$,~$T_{\rm max}(a)$] of observed time intervals
(see the caption to Fig.\hbox{}~\ref{Fig.periods}) is independent
of $c_0$. With such weak dependence on the speed of sound in the
disc, we have only two matching parameters: the specific spin $a$
and the mass $M$ of the BH.

In the flare precursor section we associate the period $T_{\rm
ms}$ with the $2200 \pm 300$ s one (group 5, \cf\ Table 2 in
\cite{asc04}) and the time interval $T_1$ with the period of $1100
\pm 100$ s (group 4 in~\cite{asc04}). In consistency with the
infra-red observations of the flare, the periods $T_1$, $T_2$,
etc.\ chirp with the peak number~\cite{gen03}, \ie\ resemble the
QPO structure and thus form a cumulative peak of a larger width
shifted to higher frequencies on the flares' power density spectra
($700 \pm 100$ s, group 3, \cf\ Fig.\hbox{}~3a and 4a
in~\cite{asc04}). We can estimate the average frequency of this
peak as $1/\overline T = 1/2 \left( 1/T_{\rm min} + 1/T_{\rm
max}\right)$. The results of the periods' matching procedure are
shown in Fig.\hbox{}~\ref{Fig.aM}. Despite large uncertainties in
the observational data allowing significant freedom of $a$ and
$M$, high positive values of $a$ (\ie\ the disc orbiting in the
same direction as the BH spin) are clearly ruled out.

\section*{Acknowledgements}
We thank the organizing committee of the Workshop for hospitality
and creating a wonderful atmosphere. We thank A.V.~Gurevich for
his interest in the work and for his support, useful discussions
and encouragement. We are grateful to K.A.~Postnov for his help
and inspiring suggestions regarding the observational part. This
work was supported by the Russian Foundation for Basic Research
(grant no.~1603.2003.2), Dynasty fund, and ICFPM.


\begin{thebibliography}{0}
\bibitem{sha83}
Shapiro S.L., Teukolsky S.A., \textit{Black Holes, White Dwarfs,
and Neutron Stars} (Wiley--Interscience Publication, New York,
1983).
\bibitem{lip92}
Lipunov V.M., \textit{Astrophysics of Neutron Stars}
(Springer-Verlag, Berlin, 1992).
\bibitem{lyn69}
Lynden-Bell D., \textit{Nature} \textbf{223}, 690 (1969).
\bibitem{sha72}
Shakura N.I., AZh \textbf{49}, 921 (1972).
\bibitem{sha73}
Shakura N.I., Sunyaev R.A., \textit{A\&A} \textbf{24}, 337 (1973).
\bibitem{nov73}
Novikov I.D., Thorne K.S. \textit{Black Holes} (C. DeWitt, B.
DeWitt., eds, Gordon and Breach, New York, 1973).
\bibitem{pac81}
Paczy\'nski B., Bisnovatyi-Kogan G.S., \textit{Acta Astron.}
\textbf{31}, 283 (1981).
\bibitem{abr88}
Abramowicz M.A., Czerny B., Lasota J.-P., Szuszkiewicz E.,
\textit{ApJ} \textbf{332}, 646 (1988).
\bibitem{pap94}
Papaloizou J., Szuszkiewicz E., \textit{MNRAS} \textbf{268}, 29
(1994).
\bibitem{rif95}
Riffert H., Herold H., \textit{ApJ} \textbf{450}, 508 (1995).
\bibitem{che97}
Chen X., Abramowicz M.A., Lasota J.-P., \textit{ApJ} \textbf{476},
61 (1997).
\bibitem{nar97}
Narayan R., Kato S., Honma F., \textit{ApJ} \textbf{476}, 49
(1997).
\bibitem{pei97}
Peitz J., Appl S., \textit{MNRAS} \textbf{286}, 681 (1997).
\bibitem{bel98}
Beloborodov A.M., \textit{MNRAS} \textbf{297}, 739 (1998).
\bibitem{gam98a}
Gammie C.F., Popham R., \textit{ApJ} \textbf{498}, 313 (1998).
\bibitem{gam98b}
Gammie C.F., Popham R., \textit{ApJ} \textbf{504}, 419 (1998).
\bibitem{art01} Artemova Yu.V., Bisnovatyi-Kogan
G.S., Igumenshchev I.V., Novikov I.D., \textit{ApJ} \textbf{549},
1050 (2001).
\bibitem{igu97}
Igumenshchev I.V., Beloborodov A.M., \textit{MNRAS} \textbf{284},
767 (1997).
\bibitem{bal98}
Balbus S.A., Hawley J.F., \textit{Rev.\hbox{} Mod.\hbox{}
Phys.\hbox{}} \textbf{70}, 1 (1998).
\bibitem{kro02}
Krolik J.H., Hawley J.F., \textit{ApJ} \textbf{573}, 754  (2002).
\bibitem{abr97}
Abramowicz M.A., Lanza A., Percival M.J., \textit{ApJ}
\textbf{479}, 179 (1997).
\bibitem{bes04}
Beskin V.S., \textit{this volume} (2004).
\bibitem{abr81}
Abramowicz M.A., Zurek W., \textit{ApJ} \textbf{246}, 314 (1981).
\bibitem{cha96}
Chakrabarti S., \textit{ApJ} \textbf{471}, 237 (1996).
\bibitem{lan87a}
Landau L.D., Lifshits E.M., \textit{The Classical Theory of
Fields} (4th edn. Butterworth-Heinemann, 1987).
\bibitem{bes97}
Beskin V.S., \textit{Phys.\ Usp.} \textbf{40}, 659 (1997).
\bibitem{bes02}
Beskin V.S., Kompaneetz R.Yu., Tchekhovskoy A.D., \textit{Astron.\
Lett.} \textbf{28}, 543 (2002).
\bibitem{lan87b}
Landau L.D., Lifshits E.M., \textit{Fluid mechanics} (2nd edn.
Butterworth-Heinemann, 1987).
\bibitem{fro98}
Frolov V.P., Novikov I.D., \textit{Black Hole Physics} (Kluwer
Academic Publishers, Dordrecht, 1998).
\bibitem{asc04}
Ashenbach B., Grosso N., D. Porquet, and Predehl P.,
\textit{A\&A}, accepted, \hbox{astro-ph/0401589} (2004).
\bibitem{gen03}
Genzel R., Sch\"odel R., Ott T. \etal, \textit{Nature}
\textbf{425}, 934 (2003).
\bibitem{car68}
Carter B., \textit{Phys.\ Rev.} \textbf{174}, 5 (1968).
\bibitem{bes93}
Beskin V.S., Pariev V.I., \textit{Phys.\ Usp.} \textbf{36}, 529
(1993).
\bibitem{bon52}
Bondi H., \textit{MNRAS} \textbf{112}, 195 (1952).
\bibitem{igu00} %
Igumenshchev I.V., Abramowicz M.A., Narayan R., \textit{ApJ}
\textbf{537}, L27 (2000).
\bibitem{pac80}
Paczy\'nski B., Wiita P.J., \textit{A\&A} \textbf{88}, 23 (1980).
\bibitem{tho86}
Thorne K.S., Price R.N., Macdonald D.A., \textit{Black Holes: The
Membrane Paradigm} (Yale Univ. Press, New Haven, CT, 1986).
\end{thebibliography}
\end{document}